\documentstyle[aps,preprint]{revtex}

\begin{document}
\draft

\title{
Measurement of Pion Enhancement at Low Transverse Momentum and of the
$\Delta$ Resonance Abundance in Si-Nucleus Collisions at AGS Energy
}

\author{ J. Barrette$^3$, R.  Bellwied$^8$, P.  Braun-Munzinger$^6$,
W. E.  Cleland$^5$, T. M. Cormier$^8$, G. David$^6$, J. Dee$^6$, G. E.
Diebold$^9$, O. Dietzsch$^7$, J. V. Germani$^9$, S. Gilbert$^3$, S. V.
Greene$^9$, J. R. Hall$^4$, T. K.  Hemmick$^6$, N. Herrmann$^2$, B.
Hong$^6$, K. Jayananda$^5$, D. Kraus$^5$, B. S. Kumar$^9$, R.
Lacasse$^3$, D. Lissauer$^1$, W. J. Llope$^6$, T. W.  Ludlam$^1$, S.
McCorkle$^1$, R. Majka$^9$, S. K. Mark$^3$, J. T.  Mitchell$^9$, M.
Muthuswamy$^6$, E. O'Brien$^1$, C. Pruneau$^3$, M. N.  Rao$^6$, F.
Rotondo$^9$, N. C. daSilva$^7$, U.  Sonnadara$^5$,
J.  Stachel$^6$, H. Takai$^1$, E. M. Takagui$^5$, T. G.  Throwe$^1$,
G. Wang$^3$, D. Wolfe$^4$, C. L. Woody$^1$, N. Xu$^6$, Y.  Zhang$^6$,
Z. Zhang$^5$, C. Zou$^6$\\ (E814 Collaboration) }

\address{
$^1$ Brookhaven National Laboratory, Upton, NY 11973}
\address{
$^2$ Gesellschaft f\"ur Schwerionenforschung, Darmstadt, Germany}
\address{
$^3$ McGill University, Montreal, Canada}
\address{
$^4$ University of New Mexico, Albuquerque, NM 87131}
\address{
$^5$ University of Pittsburgh, Pittsburgh, PA 15260}
\address{
$^6$ SUNY, Stony Brook, NY 11794}
\address{
$^7$ University of S\~ao Paulo, Brazil}
\address{
$^8$ Wayne State University, Detroit, MI 48202}
\address{
$^9$ Yale University, New Haven, CT 06511}

\date{\today}
\maketitle

\begin{abstract}
We present measurements of the pion transverse momentum ($p_t$) spectra
in central Si-nucleus collisions in the rapidity range $2.0 < y < 5.0$
for $p_t$ down to and including $p_t$ = 0.  The data exhibit an enhanced
pion yield at low $p_t$ compared to what is expected for a purely
thermal spectral shape.  This enhancement is used to determine the
$\Delta$ resonance abundance at freeze-out. The results are consistent
with a direct measurement of the $\Delta$ resonance yield by reconstruction of
proton-pion pairs and imply a temperature of the system at freeze-out
close to 140 MeV.

\end{abstract}


\narrowtext

Collisions of heavy nuclei at ultrarelativistic energies produce a zone
of hot, compressed matter.  Information from measurements of transverse
energy production \cite{814prl1,814prl2} and baryon distributions
\cite{814zphys1,802prl1,gonin} indicate that baryon densities up to ten
times normal nuclear matter density are reached during the collision
\cite{landau1,rqmd1,arc1}. This highly compressed system then expands
\cite{814plb1} until its constituents cease to interact, {\it i.e.}
``freeze out''. The expansion is reflected in the slopes of transverse
momentum spectra at midrapidity, which systematically become flatter
with increasing particle mass \cite{jsgy,jsqm93}. Very recently
\cite{814prl3} sidewards flow was directly identified for Au+Au
collisions at AGS energy. Such flow effects demonstrate
the presence of large thermal pressures, and should provide information
on the equation of state of the hot and dense matter formed in the
collision. At the same time, the connection between the transverse
momentum spectra of hadrons and the temperature of the fireball at
freeze-out is complicated by additional parameters such as flow
velocities and flow profile.

To provide information on the composition of the fireball formed in the
collision, and to get an independent measurement of the freeze-out
temperature we report here measurements of the double differential cross
sections for charged pions near the beam rapidity ($y$ = 3.4) in central
${\rm^{28}Si}$+Al, Pb collisions at ${p_{lab}}$ = 14.6 GeV/c per
nucleon.  The aim of these experiments was to measure transverse
momentum ($p_t$) spectra for pions in a kinematical region where
predicted enhancements at low $p_t$ would show up strongly. The
experimental status of low $p_t$ phenomena in high energy nuclear
collisions and possible interpretations have been summarized recently
\cite{Gillo}. At AGS energies a major source of the enhancement is
expected to be \cite{stock,Johanna} the pions produced by the decay of
the $\Delta(1232)$ resonance. The decay preferentially populates the
spectrum at low $p_t$ wherefrom the abundance of the $\Delta$ at
freeze-out can be inferred. This can then be used \cite{Johanna} to
determine the system's true temperature. To provide further support for
the feeding scenario and independent information on the $\Delta$
abundance, we also present the results of an effort to directly
reconstruct the $\Delta^{++}$ using the p$\pi^+$ invariant mass
spectrum.

The experiment was performed using the E814 apparatus at the AGS at
Brookhaven National Laboratory.  The apparatus is described in detail
elsewhere \cite{814prl1,814prl2,814zphys1,JD}.  A 14.6 GeV/c per nucleon
$^{28}\rm{Si}$ beam was incident upon Pb targets of thicknesses of 1.1
and 2.2 g/cm$^2$ and Al targets of 0.33 and 0.66 g/cm$^2$, corresponding
to 1.2 and 2.4 \% of a silicon interaction length, respectively.
Collision centrality was determined via a charged particle multiplicity
measurement in the interval ${\rm 0.85 < \eta < 3.8}$. Experimental
details and the connection between centrality and charged particle
multiplicity are discussed in \cite{jhall}.

Particles emitted in the forward direction were accepted and analyzed
by a forward spectrometer.  We define z along the incident beam, y
vertically upward and x so as to make a right handed coordinate
system.  The spectrometer aperture $-115 {\rm mr} < \theta _{\rm x} <
14 {\rm mr}$ and $|\theta _{\rm y}| < 21 {\rm mr}$ was defined by a
Pb/steel collimator.  Accepted particles pass through a dipole magnet
and are momentum analyzed via a pair of drift/pad tracking
chambers. The momentum resolution of the spectrometer has been
modelled using the GEANT package \cite{geant}.  At the low field used
in the present measurements, the resolution in momentum {\it p} is
dominated by multiple scattering and is nearly uniform in momentum
with $\delta p/p \sim 4.1 \%$ for the momentum range considered here.
Scattering in the target creates a distortion in $p_t$ without
significantly altering {\it p}.  This effect together with all other
imperfections implies that $\delta p_t < 4$ MeV/c for $p_t < 100$
MeV/c and $\delta p_t/p_t =4\%$ for larger $p_t$ values. All data are
presented in 10 MeV/c $p_t$ bins.

Time-of-Flight and hence velocity is determined by one of two
scintillator hodoscopes located 12 m (200 psec resolution) and 31 m (350
psec resolution) from the target.  The spectrometer is capable of
separating protons and pions up to $p$ = 7 GeV/c.  Background in the
pion sample due to kaons and unrecognized decays is less than 10\%.
Possible electron contamination was investigated in two ways.  First,
for momenta below 0.5 GeV/c electrons can be separated via
time-of-flight. In this momentum range, the electron to pion ratio is
observed to decrease with increasing {\it p} and is close to 5\% at 0.5
GeV/c.  Simulations imply that the primary source of electron
contamination is photon conversion in the target.  A comparison of
results for the 1\% and 2\% targets shows no statistically significant
evidence for electron contamination in the pion sample presented
here. We conclude that electron contamination can be
neglected. Consequently, the data obtained with the 1.1 and 2.2 \%
targets were combined.

Fig. ~\ref{Fig:mt} shows a summary of the measured $\pi^-$ transverse
mass ($m_t = \sqrt{p_t^2+m_{\pi}^2}$) spectra for central
($\sigma/\sigma_{geo} \leq 2\%$) Si+Pb collisions. Here $m_{\pi}$ is the
pion rest mass. The vertical axis is $1/m_t^2 \times d^2N/dm_tdy$, the
representation in which a Boltzmann (or thermal) distribution is a pure
exponential in $m_t$.  Our acceptance in $m_t$ is largest for rapidities
$3.0 < y < 4.0$.  The measurable range in $m_t$ is limited by the
geometrical opening of the spectrometer at low {\it y} and particle
identification at high {\it y}.  The solid curves in Fig. ~\ref{Fig:mt}
show Boltzmann fits to the data, with the fit interval restricted to
($m_t$~-~$m_\pi$~$)>$~$160$ MeV/c$^2$. The data exhibit a significant
enhancement over this functional form at low $m_t$.  This is most
clearly seen in the inset which shows the ratio of the data to the
thermal fit on a linear scale.  The dashed line demonstrates that the
data rise also faster than expected from an invariant distribution
$1/m_t \times d^2N/dm_tdy \propto \exp(-m_t/T)$. Results for $\pi^+$, and
for the Al target, (not shown here) exhibit enhancements of similar
strength as those in Fig. ~\ref{Fig:mt}.  The strength of the
enhancement does not vary significantly with centrality in the range
$\sigma/\sigma_{geo}$ = (2-10)\%.

The large stopping at AGS energies \cite{814zphys1} implies that the
fireball formed in the collision is baryon-rich.  Therefore, baryonic
resonances and in particular the $\Delta(1232)$, in the following
simply denoted $\Delta$, with its low decay momentum of 227 MeV/c are
therefore anticipated to be the major contributors to pion spectra at
low $p_t$ \cite{Johanna,Heinz}.  One can extract from the pion
spectra an estimate of the fraction of pions resulting from $\Delta$
decay.  To do this we have computed the pion spectral shape by
superposition of direct thermal pions (with spectral slope adjusted at
high $m_t$) and pions from $\Delta$ decay for various ratios $f =
\pi_{\Delta}/\pi_{direct}$.  Rapidity and $p_t$ distributions of the
$\Delta$ were taken to follow predictions by the RQMD model
\cite{rqmd2} but the assumption that the $\Delta$ distributions follow
those of the measured \cite{814zphys1} protons yields very similar
results. Fig. ~\ref{Fig:ratios} shows the ratio of the pion spectra to
the fitted Boltzmann distribution.  The data are compared to model
calculations using {\it f}=0.4 and 0.6.  The calculations bracket the
data quite well and establish that the ratio $f = 0.5 \pm 0.1$
implying that 1/3 of all pions come from $\Delta$ decay without
further interaction.  The ratio between the number of pions and
nucleons for central Si+Pb collisions is observed to be $(\pi/N)_{exp}
\approx 1.07$ \cite{jsqm93,bodrum}. Hence the fraction of nucleons
excited to the $\Delta$ resonance at freezeout is $0.36 \pm 0.05$.

The measured pion spectra are also well reproduced by cascade models
such as RQMD\cite{rqmd2} and ARC\cite{arc1} which have the $\Delta$
resonance explicitely built into the collision dynamics.  This is
illustrated by the dashed lines in Fig. ~\ref{Fig:mt} representing the
RQMD prediction, which accounts for shape and absolute yield of the data
once the experimental trigger conditions are incorporated.  Predictions
using the RQMD and ARC models are in good agreement with all our pion data (for
both charges and both targets). The dominant source of the rise at low
$m_t$ in pion spectra calculated with RQMD can be traced back
\cite{rqmd2} to the $\Delta(1232)$ resonance decay. The overall
predicted freeze-out $\Delta$ excitation probability of 0.35 is very
close to the experimental value given above.

Close inspection of Fig. ~\ref{Fig:mt} reveals that in addition to the
low $p_t$ enhancement discussed above there is also visible in the data
an increase with even steeper slope at very low transverse momenta ($p_t
< 50$ MeV/c corresponding to $m_t-m_{\pi} < 0.01$ GeV/c$^2$). A similar
effect is seen \cite{corm} in our recent measurement covering backward
rapidities with a different detector. Whether this is also due to
resonance decays (such as $\eta$ decay, see \cite{Johanna}) or has a
more exotic origin (such as chiral symmetry restoration \cite{gerry})
remains to be quantitatively explored.

To get an independent measurement of the number of $\Delta$'s at
freeze-out in Si+Pb collisions we have reconstructed the $\Delta^{++}$
via its decay to p$\pi^+$.  This is the most easily measured of all
$\Delta$ decays since (i) its branching ratio is nearly 100\%, (ii) all
particles in the final state are charged, and (iii) there is no
interference from $\Lambda$ decay (which could disturb
$\Delta^0$ measurements).  Additionally, the asymmetry of the E814
spectrometer makes it best suited for like-sign pair measurements.

The invariant mass for p$\pi^+$ pairs was reconstructed for central
($\sigma/\sigma_{geo} \leq 10\%$) collisions.  Protons near beam
rapidity (y~$\geq$~3.1) were rejected from the sample since they are in
part projectile fragments.  Pions below rapidity 3.0 were rejected since
the $\Delta$ decay kinematics does not permit such pions into the E814
spectrometer acceptance with the proton rapidity cut used.

Fig. ~\ref{Fig:delta} shows a summary of our measurement of the
$\Delta^{++}$ using p$\pi^+$ pairs.  The analysis employs the ``mixed
events'' technique.  This method allows to separate a small signal from
a large combinatorial due to uncorrelated pairs.  One determines the
shape of the combinatorial background by constructing an invariant mass
spectrum using (uncorrelated) protons and pions from different
events. The resulting distribution is normalized (see below) to the true
pair spectrum and subtraction yields the signal.  To normalize, we treat
the combinatorial spectrum as a function of a single free parameter, the
normalization constant.  This function is fitted to the high mass end
($M_{inv} > M_1$) of the true pair spectrum, choosing a normalization
constant which minimizes $\chi^2$.  The net $\Delta^{++}$ yield was
found to not differ beyond statistics using $M_1$ values in the range
$1.4<M_1< 1.8$ GeV/c$^2$.  Lower values of $M_1$ result in
over-subtraction and higher values suffer from statistical
uncertainties. Note that the true pair acceptance for decay of a nuclear
resonance with M $\geq$ 1.4 GeV/c$^2$ is vanishing.  A value of $M_1
=1.4 {\rm GeV/c^2}$ was selected for the analysis.  The solid and dashed
histograms in Fig. ~\ref{Fig:delta} show the results for the true pair
invariant mass ($M_{inv}$) distribution and the normalized combinatorial
background distribution.

The robustness of the analysis procedure was tested via Monte Carlo
analysis.  A ``negative test'' data sample of uncorrelated p$\pi^+$
pairs was generated using the measured single particle distributions.
Analysis of these data shows no $\Delta$ signal.  Additionally, a
``positive test'' was performed on p$\pi^+$ pairs from RQMD-generated
events to verify that the $\Delta$ resonance is observed correctly
there.

The $M_{inv}$ spectrum of p$\pi^+$ pairs after background subtraction is
shown in the bottom panel of Fig. ~\ref{Fig:delta}.  The total yield
into the E814 acceptance for $ 4.01 \times 10^5$ central collisions is
587 $\pm$ 165 $\Delta^{++}$.  We have computed the acceptance of the
E814 spectrometer for $\Delta^{++}$ using GEANT. Over the rapidity
interval $1.9 < y < 3.1$ covered by the experiment the acceptance varies
between $3 \times 10^{-4}$ at y=2 and $1.2 \times 10^{-2}$ at
y=3. Assuming that the shape of the $\Delta$ rapidity and $p_t$
distribution is close to that measured for protons \cite{814zphys1,JD},
and taking into account a track reconstruction efficiency of 73\% per
particle, the yield corresponds to 1.7 $\pm$ 0.5 $\Delta^{++}$ per
central Si+Pb collision into $1.9 < y < 3.1$. Using the $\Delta$
distributions in {\it y} and $p_t$ from RQMD gives a very similar
result.

The predicted $\Delta^{++}$ yield at freeze-out from the RQMD model is
in good agreement with the measurement. To best approximate our
experimental conditions, we have chosen to analyze RQMD events in a
similar manner as the actual data and extract the predicted $\Delta$
yield using a combinatorial mass spectrum and a mixed-event
subtraction. The RQMD prediction is then 1.6 $\pm$ 0.3 $\Delta^{++}$ per
event in the rapidity interval $1.9 < y < 3.1$, in remarkable agreement
with our measurement. The same calculation yields a rapidity integrated
yield of $\Delta^{++}$ of 14.7 $\pm 0.9$ and finds 35 \% of all nucleons in the
$\Delta$ resonance at freeze-out.

Assuming thermal equilibrium, the measured $\Delta$ abundance yields
information \cite{Johanna} on the freeze-out temperature of the system,
independent of analyses of the slopes of particle spectra. To determine
the freeze-out temperature {\it T}, we calculated the number density of
all nonstrange baryonic resonances with masses less than 2 GeV/c$^2$ as
a function of temperature.  The calculation closely follows that of
\cite{Johanna} but takes, in addition, into account the widths of all
states.  The results for population ratios (which are essentially
independent of baryon chemical potential) are presented in
Fig. ~\ref{Fig:tempdep} as a function of {\it T} for all included
baryons.  Using the population ratio for $\Delta$'s of 0.36$\pm 0.05$
determined from our data we extract the freeze-out temperature to be $T
= 138 ^{+23}_{-18}$ MeV.

In summary, we have shown that pion spectra from Si-nucleus collisions
at AGS energy exhibit a significant enhancement at low $p_t $.  A simple
model incorporating pions from $\Delta$ decay accounts quite accurately
for the observed shape if $\Delta$/nucleon ratios in the range 0.36
$\pm$ 0.05 are assumed.  RQMD calculations are consistent with our
result and also reproduce accurately the measured pion spectra.
Additionally, the results from the analysis of spectral shapes are
consistent with a direct measurement of the $\Delta^{++}$ abundance for
Si+Pb, which yields 1.7 $\pm$ 0.5 $\Delta^{++}$ per collision in the
rapidity interval $1.9 < y < 3.1$.  Our measurement thus quantitatively
establishes the importance of the $\Delta$ resonance to the dynamics of
the collision.  We find that the same concentration of freeze-out
$\Delta$'s simultaneously explains our pion enhancement and the directly
measured yield of $\Delta^{++}$.  Finally, the results imply that in
central Si+Pb collisions a fireball is formed with substantial
excitation of $\Delta$ baryons which freezes out at $T = 138
^{+23}_{-18}$ MeV.

We wish to thank the Brookhaven Tandem and AGS staff for their excellent
support and are particularly grateful for the expert help of W.
McGahern and Dr. H. Brown. R. Hutter and J. Sondericker provided
important technical support.  Financial support by the US DoE, the NSF,
the Canadian NSERC, and CNPq Brazil is gratefully acknowledged.

\begin{figure}
\caption{Pion transverse mass spectra for central
($\sigma/\sigma_{geo}=2\%$) Si + Pb collisions in different rapidity
intervals.  Starting with {\it y}=4.7, the distributions in each successively
lower {\it y} bin have been multiplied by increasing powers of 10.  Shown in
the inset are the ratios: Data/Boltzmann (points) and $m_t$
exponential/Boltzmann (dotted line). For more details see
text. \label{Fig:mt}}
\end{figure}

\begin{figure}
\caption{Pion transverse mass spectra plotted as a ratio of the data
to the best fit Boltzmann distribution ($m_t \geq 300 {\rm MeV/c^2}$).
Also shown are predictions from a model containing direct pions and
$\Delta$ decay pions in the ratios $\pi_{\Delta}/\pi_{direct}$ = 0.4 and
0.6 (lower and upper curves respectively). \label{Fig:ratios}}
\end{figure}

\begin{figure}
\caption{Reconstruction of the $\Delta^{++}$ resonance.  Shown in the top
panel are the invariant mass spectra of p$\pi^+$ pairs from the same
event (solid histogram) and of p$\pi^+$ pairs from mixed events (dashed
histogram).  The bottom panel shows the difference in the true pair and
mixed event pair spectra in which the $\Delta$ resonance is
visible.\label{Fig:delta}}
\end{figure}

\begin{figure}
\caption{Thermal occupation probabilities of non-strange baryons with
mass $ m <$ 2 GeV/c$^2$ as function of temperature. For details see
text. \label{Fig:tempdep}}
\end{figure}

\end{document}